\documentclass[epjCONF,columns]{svjour} 
\usepackage{graphics}
\usepackage[varg]{txfonts} 
\usepackage[latin1]{inputenc}
\session-title{FUSION11}
\begin{document}
\title{Phase diagram of dilute cosmic matter}
\author{Yoritaka Iwata}
\institute{GSI Helmholtzzentrum f\"ur Schwerionenforschung, Darmstadt, Germany}
\abstract{
Enhancement of nuclear pasta formation due to multi-nucleus simultaneous collision is presented based on time-dependent density functional calculations with periodic boundary condition. 
This calculation corresponds to the situation with density lower than the known low-density existence limit of the nuclear pasta phase.  
In order to evaluate the contribution from three-nucleus simultaneous collisions inside the cosmic matter, the possibility of multi-nucleus simultaneous collisions is examined by a systematic Monte-Carlo calculation, and the mean free path of a nucleus is obtained. 
Consequently the low-density existence limit of the nuclear pasta phase is formed to be lower than believed up to now.
} 
\maketitle
\section{Introduction}
Nuclear pasta, whose existence and role were suggested in association with the collapse driven supernova explosions \cite{watanabe,sowa}, is of great importance with respect to the astrophysical synthesis of chemical elements.
Indeed nuclear pasta is known to change the transparency rate of neutrinos \cite{ravenhall,sato}, which play a role of trigger in supernova explosions giving rise to a wide variety of chemical elements.
Therefore the existence of nuclear pasta at low densities is expected to have a great impact on clarifying the historical evolution of chemical elements and also on revising the low-density and low-temperature part of the nuclear phase diagram \cite{sonoda,lamb}.
In particular, here we are concerned with its existence in the outermost part of protoneutron stars (for the review of supernova mechanisms, see \cite{bethe}).

In this paper, based on three-dimensional time-dependent density functional calculations, a dynamical formation of nuclear pasta in the low-density and low-temperature situations is demonstrated by 3-nucleus simultaneous heavy-ion collision. 
This calculation simulates the situation outside of the known low-density existence limit of nuclear pasta phase.  
In order to examine the validity of this calculation, the possibility of multi-nucleus simultaneous collisions is studied by a systematic three-dimensional Monte-Carlo calculation, and the mean free path of a nucleus at a given density and temperature is obtained. 
Note that for this kind of nuclear pasta formation, the presence or absence of the fast charge equilibration process \cite{iwata} determines the upper boundary of nuclear pasta phase with respect to the temperature \cite{iwata2}.

\section{Heavy-ion collisions in cosmic matter}
\subsection{Dilute cosmic matter}
Cosmic nuclear matter, which consists of both neutrons and protons, is studied.
Here we are concerned with the dilute cosmic matter whose density is less than 10 \% of the standard nuclear density $\rho_0$.
Nuclear matter of this kind is expected to exist in the outermost part of the protoneutron star.
In the following the dilute cosmic matter means the matter whose density is less than 10~\% of the standard nuclear density. 

Although the existence of nuclear pasta was suggested in more dense situations, its existence on the outermost part of protoneutron star has never been understood. 
Its existence, however, is shown in this paper.

\begin{figure*}
\resizebox{2.00\columnwidth}{!}{%
\includegraphics{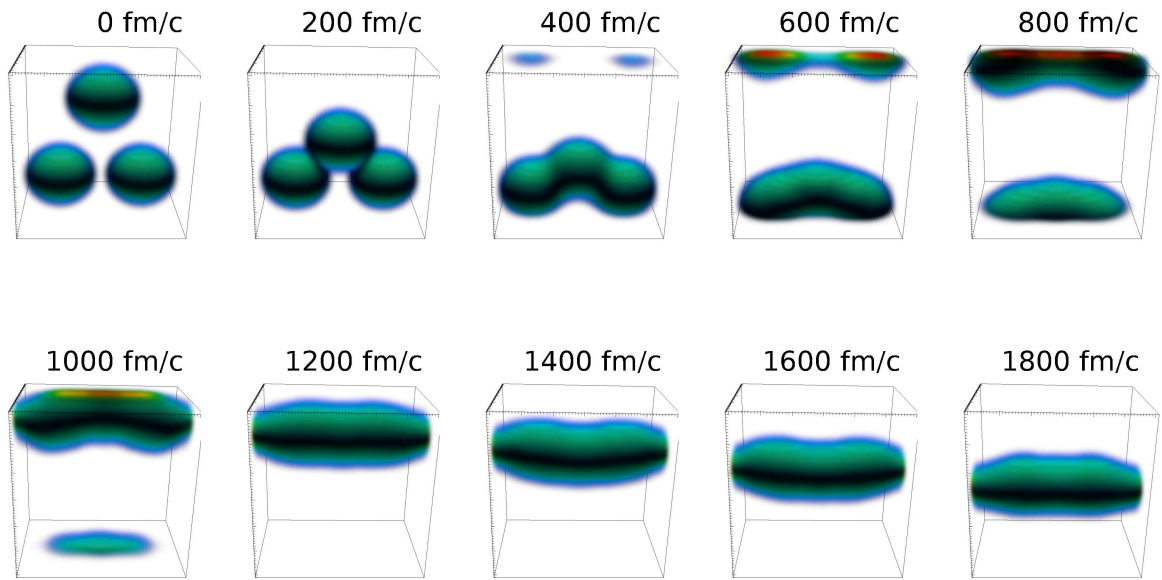} }
\caption{\label{fig1} (color online) Time evolution of a 3-nucleus simultaneous collision in a cubic space with periodic boundary condition. 
The box size is 64~fm $\times$  32~fm  $\times$ 64~fm. }
\end{figure*}

\subsection{Multiple nucleus collisions}
We are concerned with heavy-ion collisions appearing in the dilute cosmic matter.
Here the mean free path of a nucleus is important, and multi-nucleus simultaneous collisions might play a role.
Indeed, rather dense situations, which are not achieved on the earth, appear in the universe, so that the presence of multi-nucleus simultaneous collisions is expected.
For reference, the density of targets in accelerators is roughly estimated by
\[  \rho_{\rm Lab} = 10^{-12} \rho_0 \]
where the radii of nuclei and atoms are assumed to be 10$^{-14}$ m and 10$^{-10}$ m, respectively.
In the following the density is denoted by $\rho$, and the terminology of multi-nucleus simultaneous collision exactly means simultaneous collisions between more than two nuclei. 

\section{Calculations of 3-nucleus simultaneous collisions}
A three-nucleus simultaneous collision is shown based on three-dimensional time-dependent density functional calculations employing the Skyrme force parameter set SLy6 \cite{Chabanat-Bonche}.
The calculation is carried out in a spatial box $64 \times 32 \times 64$ fm$^3$ with a spatial grid spacing of 1.0 fm, where periodic boundary condition is applied.
Three $^{78}$Zn are initially located at (8,0,-8), (-8,0,-8) and (0,0,8), respectively (cf. the first frame in Fig. \ref{fig1}).
The two nuclei initially located at (8,0,-8) and (-8,0,-8) are not given velocities in the center-of-mass frame, while the nucleus initially located at (0,0,8) has the initial velocity along the z-axis.
Accordingly, the total system consists of 90 protons and 144 neutrons, and the total incident energy is set to 100~MeV in the center-of-mass frame.

Figure~\ref{fig1} demonstrates a 3-nucleus simultaneous collision resulting in nuclear pasta.
The temperature and density of this calculation are 0.5~MeV (see following discussion leading to Eq. (\ref{sb})) and 0.039 $\rho_0$, respectively.
This situation corresponds to the state with a density lower than the known low-density existence limit of nuclear pasta phase \cite{sonoda}.
If we consider collisions with a higher energy ($T =$ 7.5~MeV), nuclear pasta cannot be formed.
It corresponds to the energy above the upper-limit energy of fast charge equilibration \cite{iwata}.
For comparison, 2-nucleus collisions require more dense situations to form a nuclear pasta.  
A calculation with exactly the same setting except for the force parameter set (SKI3 \cite{reinhard}) results in the same conclusion.
Consequently the low-density existence limit of nuclear pasta phase is suggested to be extended by taking into account multi-nucleus simultaneous collision dynamics.
It is important to note that the boundary condition is not essential.
Therefore Fig.~\ref{fig1}  should not be interpreted as the formation of infinitely connected matter, but simulates the favorable situation of very long finite spaghetti-like structure.
Pasta formation by 3-nucleus simultaneous collisions has been calculated for several different settings of the initial condition \cite{iwata2}. 

\begin{figure*}
\begin{center}
\resizebox{0.80\columnwidth}{!}{%
\includegraphics{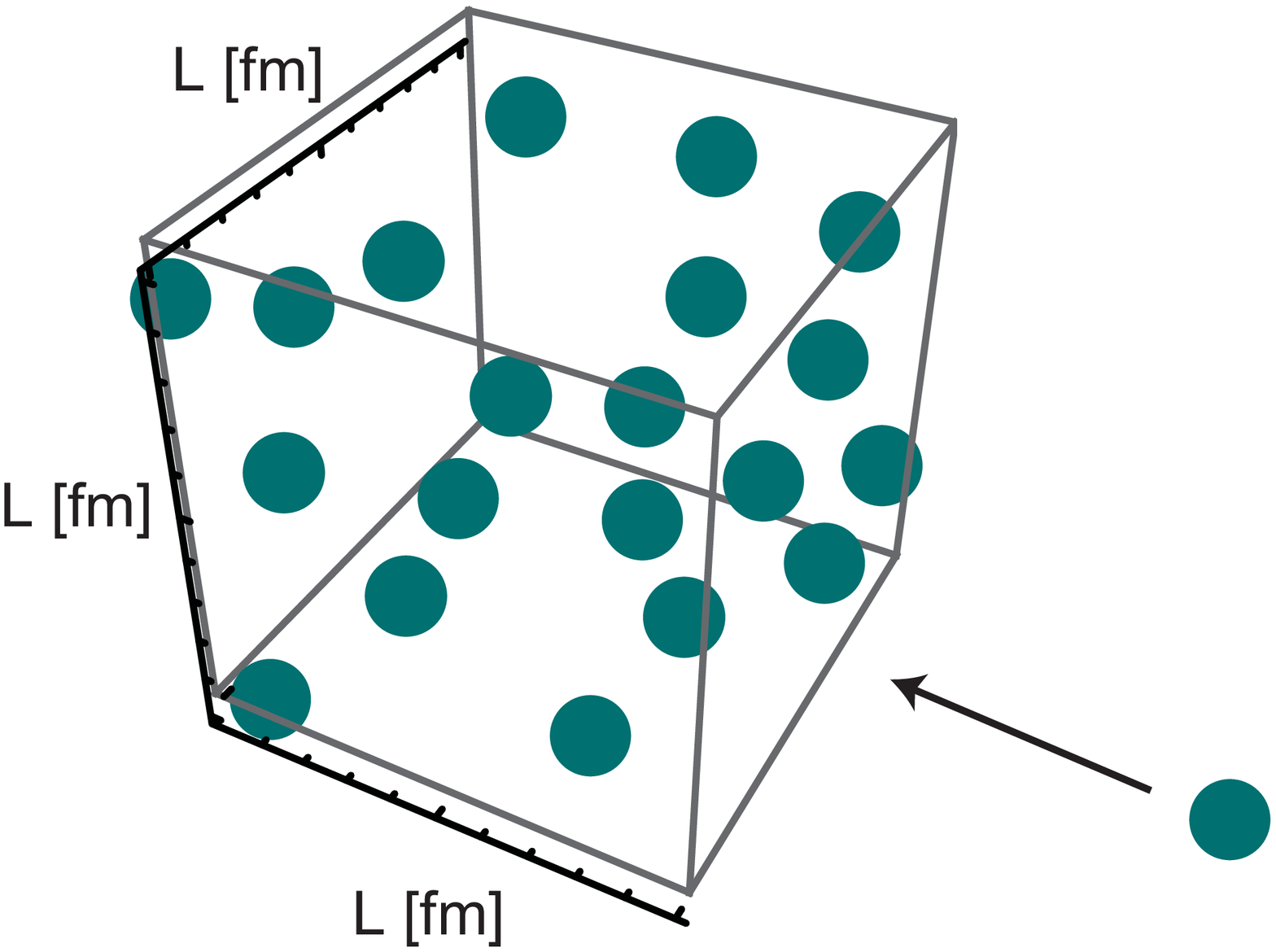}} ~ 
\resizebox{1.0\columnwidth}{!}{%
\includegraphics{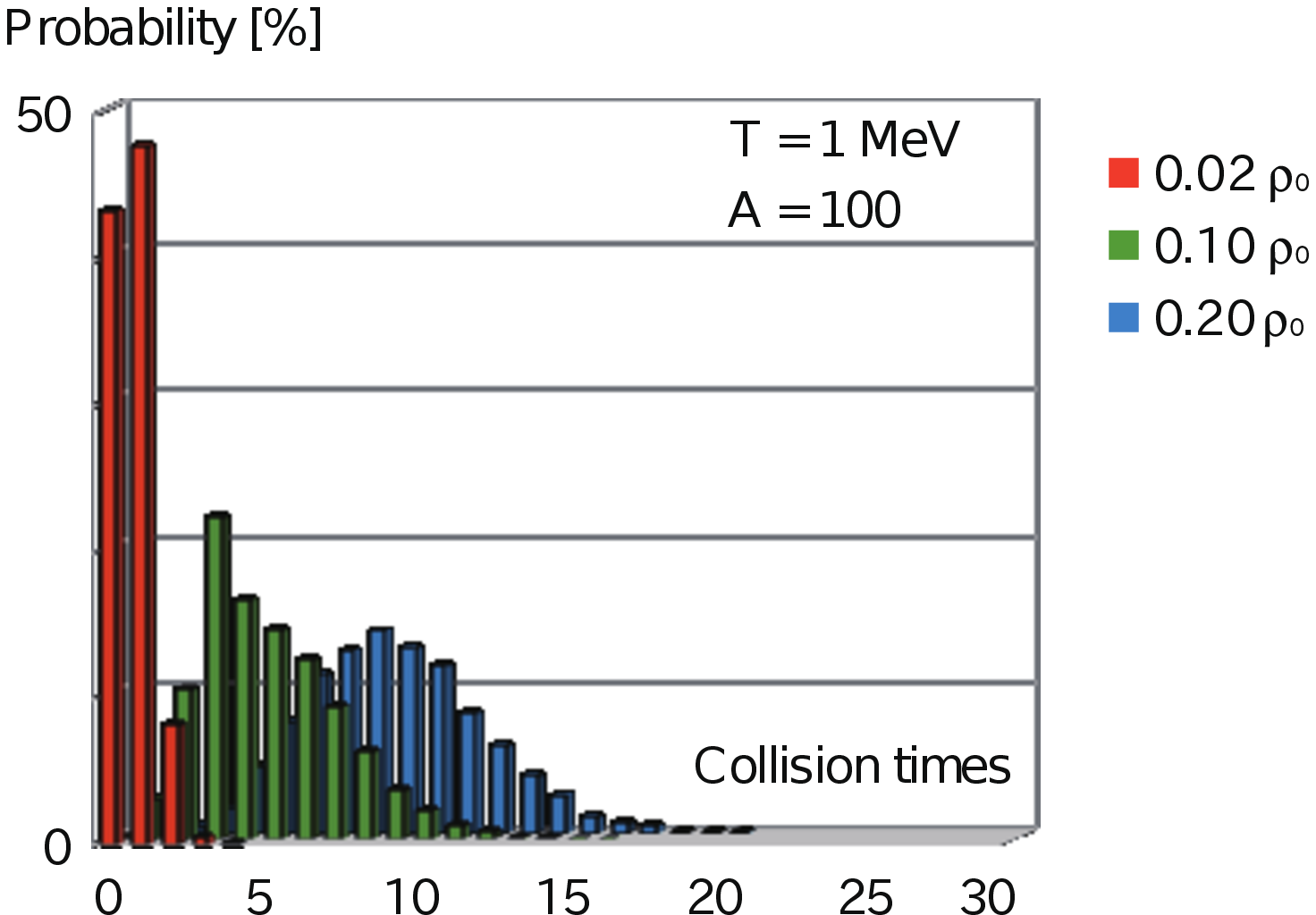}}
\caption{\label{fig2} (color online) In the left panel the setting of the Monte Carlo calculation is shown, where periodic boundary condition is applied. 
Spheres, whose spatial distribution is random, stand for nuclei.
The passing of the projectile going through nuclear matter is simulated. 
In the right panel the result of Monte Carlo calculation is shown. The number of collisions during a typical interval of low-energy heavy-ion collisions is presented.}
\end{center}
\end{figure*}

\section{Monte Carlo calculation}
Although multi-nucleus collisions are suggested to enhance nuclear pasta formation, it is worthless if such a situation is quite rare in the cosmic matter.

Let us confirm the validity of multi-nucleus collisions in the cosmic matter.
The numbers of simultaneous collisions are calculated utilizing the Monte Carlo method.
A cubic space with side length $L$~fm is given.
Spheres (corresponding to nucleus) with the same radius $r$~fm are introduced in the cubic space, where those positions are chosen randomly (The left panel of Fig.~\ref{fig2}).
Accounting for the effect from the Pauli principle (Fermionic feature) and the Coulomb force, a sphere overlapping with the other sphere is discarded from the beginning. 
Together with periodic boundary condition, thermal equilibrium of nuclear matter is simulated.
The density of a sphere is fixed to $\rho_0$, so that the density of matter is given by adjusting the values of $L$ ($L$ is fixed in actual calculations), $r$ and the number of spheres.
Another sphere (nucleus) with exactly the same feature (radius: $r$~fm, density: $\rho_0$), which corresponds to the projectile, is coming from the outside of the cube. 
The temperature of the system is determined by the relative velocities between spheres.
Here it is given by the relative velocity of the projectile to the cubic space.  
The projectile experiences collisions with spheres, when it goes though this periodic space.

Let the typical time interval of low-energy heavy-ion collisions be 1000 fm/c = $33.3 \times 10^{-22}$~s.
Collisions are defined to be simultaneous, if they occur within this time interval.
The statistical values for the number of simultaneous collisions is obtained by $10^4$ trial calculations.  
Let $A$ denote the nuclear mass number. 
Three cases with different radii of the sphere ($A=$ 50, 100, 200):
\[ r = 1.2 \times A^{1/3} ~{\rm fm}, \]
are calculated for several densities and temperatures. 
In particular, with respect to low-energy heavy-ion collisions, we consider three different temperatures $T = 1, 5, 10 ~{\rm MeV}$. 
According to the Stefan-Boltzmann relation, the temperature of matter in heavy-ion reactions is estimated by the relative kinetic energy, 
\begin{equation} \label{sb} 
\sigma T^4 = E^*, \end{equation}
where $T$ and $E^*$ mean the temperature and the incident energy, respectively.

\begin{table} 
\begin{center} 
  \caption{Mean free path [fm] of a nucleus in nuclear matter, which consists of nuclei with mass number $A$.}
\vspace{2.5mm}
(a) $A$ = 50 \\
\begin{tabular}{|c||r|r|r|r|r|r|} \hline $T~ \backslash ~ \rho$ &~ 0.02~$\rho_0$ ~&~ 0.10~$\rho_0$ ~&~ 0.20~$\rho_0$  \\ 
\hline  10 MeV  & 19.5 & 4.6  & 2.7 \\
\hline  5 MeV    & 30.9 & 6.9  & 4.1  \\
\hline 1 MeV   & 54.4 & 14.2  & 8.2  \\
\hline 
\end{tabular} \vspace{2.5mm} \\
(b) $A$ = 100 \\
\begin{tabular}{|c||r|r|r|r|r|r|} \hline $T~ \backslash ~ \rho$ &~ 0.02~$\rho_0$ ~&~ 0.10~$\rho_0$ ~&~ 0.20~$\rho_0$  \\ 
\hline  10 MeV  & 23.6 & 5.9  & 3.4 \\
\hline  5 MeV    & 40.2 & 10.4  & 5.6  \\
\hline 1 MeV   & 70.1 & 20.1  & 10.8  \\
\hline 
\end{tabular} \vspace{2.5mm} \\
(c) $A$ = 200 \\
\begin{tabular}{|c||r|r|r|r|r|r|} \hline $T~ \backslash ~ \rho$ &~ 0.02~$\rho_0$ ~&~ 0.10~$\rho_0$ ~&~ 0.20~$\rho_0$  \\ 
\hline  10 MeV  & 31.2 & 7.4  & 4.3 \\
\hline  5 MeV  & 42.5 & 11.1  & 6.4  \\
\hline 1 MeV   & 89.7 & 34.9  & 16.1  \\
\hline
\end{tabular}  \label{table1}
\end{center}
\end{table} 

\subsection{Mean free path of nucleus in the matter}

Table \ref{table1} shows the mean free path of a nucleus at given temperature and density.
The mean free path tends to be larger for matter including heavier nuclei.
It seems to be reasonable; for a fixed density and volume of the cubic space, there are more nuclei in lighter case.
The mean free path becomes smaller for dense and high temperature situations.
Comparing two cases in $\rho = 0.02 \rho_0$ and $T= 1$~MeV, almost 35~fm difference of mean free path is noticed depending only on the size of the nuclei. 
Note that the cases with $T = 1$~MeV roughly correspond to heavy-ion reactions at the incident energy slightly higher than the Coulomb barrier.
This table implies that multi-nucleus simultaneous collisions commonly appear in the dilute cosmic matter satisfying $\rho > 0.01 \rho_0$.  

\subsection{Transition to the situation dominated by multi-nucleus collisions}
The right panel of Fig.~\ref{fig2} shows the statistics of simultaneous collisions for $T = 1$~MeV and A = 100.
It demonstrates the density-dependent transition from dominance of 2-nucleus collisions to that of multi-nucleus collisions.
For the lowest density case with $\rho = 0.02 \rho_0$, the situation is dominated by 2-nucleus collisions.
However, depending on the increase of the density, 2-nucleus collisions lose their dominance below $\rho = 0.10 \rho_0$, and multi-nucleus collisions becomes dominant instead.
It is also seen that three-nucleus simultaneous collisions are not expected (their expectation values are negligibly small), if we are only interested in collisions in the laboratory ($\rho = \rho_{\rm Lab}$). 

\begin{table}  \begin{center}
  \caption{Transition density [$10^{-2} \rho_0$] from dominance of 2-nucleus collisions to that of multi-nucleus collisions.
The dominance of 2-nucleus collisions is defined to be true, if 2-nucleus collisions are the most probable among other finite-nucleus collisions (not taking into account collisionless cases).
The density dominated by the 2-nucleus collisions (the left side of arrow) and that dominated by the multi-nucleus collisions (the right side of arrow) are shown in each case. }
\vspace{2.5mm}
\begin{tabular}{|c||c|c|c|c|c|c|} \hline $T~ \backslash ~ A$ ~&~ 50 ~&~ 100 ~&~ 200  \\ 
\hline  10 MeV & 1.5 $\to$ 2.0 & 1.5 $\to$ 2.0  & 1.5 $\to$ 2.0 \\
\hline  5 MeV  & 1.5 $\to$ 2.0 & 1.5 $\to$ 2.0  & 1.5 $\to$ 2.0 \\
\hline  1 MeV  & 2.0 $\to$ 4.0 & 2.0 $\to$ 4.0  & 4.0 $\to$ 10.0 \\
\hline 
\end{tabular}  \label{table2}
\end{center} \end{table} 

Table \ref{table2} summarizes the transition from the situation dominated by 2-nucleus collisions to that dominated by multi-nucleus collisions.
The transition density is found to be of the order of $10^{-2} \rho_0$, where its temperature dependence is small.
In addition simultaneous collisions between many nuclei are expected for the nuclear matter with 1 \% of the standard density.
Indeed, for $\rho = \rho_0$ and $T= 1$~MeV, the most frequent numbers of collisions are  77, 59, 47 for $A =$ 50, 100, 200, respectively.
Note again that this calculation is carried out by giving a typical time interval (1000 fm/c), so that the transition density becomes larger if we assume the smaller time interval.
Consequently, multi-nucleus simultaneous collisions become important and are expected to play a role in the dilute cosmic matter with density higher than $0.01 \rho_0$, and the validity of the calculation shown in Fig.~\ref{fig1} is confirmed.

\section{Summary}
Enhancement of nuclear pasta formation due to 3-nucleus simultaneous collisions has been presented based on time-dependent density functional calculations (using SLy6 and SKI3 force parameter sets). 
Multi-nucleus simultaneous collisions between more than two nuclei have been confirmed to occur in cosmic matter with density higher than $0.01 \rho_0$.
Therefore, nuclear pasta formation is possible in the outermost part of protoneutron stars, and neutrino radiation is not as easy as expected.
Eventually, the low-density existence limit of nuclear pasta phase is suggested to be lower than believed up to now. \\

The author thanks to Profs. K. Iida, N. Itagaki, J. A. Maruhn and T. Otsuka.  
This work was supported by the Helmholtz Alliance HA216/EMMI.

\end{document}